\documentclass[
reprint,
aps,
superscriptaddress,
nofootinbib,
nobibnotes,
onecolumn]{revtex4-2}

\usepackage{xcolor}
\usepackage{graphicx}
\graphicspath{{./fig/}}
\usepackage{amsmath,amssymb,amsfonts,mathrsfs,bm}
\usepackage{slashed}
\usepackage[%
colorlinks=true,
linkcolor=blue,
citecolor=blue,
]{hyperref}
\usepackage{braket}

\begin{document}
\title{Enhancement of small-scale induced gravitational waves from the soliton/oscillon domination}

\author{Xiao-Bin Sui}
\email{suixiaobin21@mails.ucas.ac.cn}
\affiliation{School of Fundamental Physics and Mathematical Sciences, Hangzhou Institute for Advanced Study, University of Chinese Academy of Sciences (HIAS-UCAS), Hangzhou 310024, China}
\affiliation{CAS Key Laboratory of Theoretical Physics, Institute of Theoretical Physics, Chinese Academy of Sciences, Beijing 100190, China}
\affiliation{University of Chinese Academy of Sciences, Beijing 100049, China}

\author{Jing Liu}
\email{liujing@ucas.ac.cn}
\affiliation{International Centre for Theoretical Physics Asia-Pacific, University of Chinese Academy of Sciences, Beijing 100190, China}
\affiliation{Taiji Laboratory for Gravitational Wave Universe (Beijing/Hangzhou), University of Chinese Academy of Sciences, Beijing 100049, China}


\author{Rong-Gen Cai}
\email{cairg@itp.ac.cn}
\affiliation{Institute of Fundamental Physics and Quantum Technology, Ningbo University, Ningbo, 315211, China}

\begin{abstract}
 We investigate density fluctuations and scalar-induced gravitational waves (GWs) arising from the production of long-lived solitons and oscillons, which can dominate the early Universe and drive reheating prior to the standard radiation-dominated era.
Curvature perturbations are generated not only by the Poisson distribution of these solitons/oscillons but are also amplified during the early matter-dominated phase when the solitons/oscillons dominate. Scalar-induced GWs with characteristic energy spectra emerge from these amplified curvature perturbations, particularly at the sudden transition from matter domination to radiation domination.
We analyze this scenario and its GW predictions in detail, focusing on the impact of the initial energy fraction, the lifetime, and the average separation of these solitons/oscillons. This provides a wealth of potential observational targets for various GW detectors. Furthermore, for inflation and preheating models that produce oscillons with large separations, we derive new constraints on these models based on the upper bounds on the effective number of relativistic degrees of freedom, as anticipated from future cosmic microwave background and large-scale structure observations.
 %

\end{abstract}

\maketitle

\section{Introduction}
Gravitational waves (GWs) have been attracting more and more attention on account of
 their potential to provide valuable information about both merger events of compact binaries and the early Universe ~\cite{LIGOScientific:2016aoc,Cai:2017cbj}.
Stochastic gravitational waves backgrounds (SGWBs) induced by curvature perturbations are one of the most important cosmological GW sources tied to the evolution of the early Universe, which are expected to generate both from vacuum quantum fluctuations during inflation~\cite{Baumann:2007zm,Saito:2008jc, Ananda:2006af, Kohri:2018awv,Cai:2018dig,Cai:2019amo,Cai:2019cdl, Espinosa:2018eve, Kohri:2018awv,Domenech:2021ztg,Martin:2019nuw,Martin:2020fgl,Choudhury:2023kdb} and the violent physical processes after inflation~\cite{Khlopov:1998nm,Lozanov:2019ylm,Liu:2021svg,Liu:2022lvz,Ellis:2019oqb,Baker:2021nyl,Liu:2023tmv,Bhaumik:2022pil,Sui:2024nip,Cai:2024nln,Flores:2024lng}. Such SGWBs can be explored through multiband GW projects such as LIGO/VIRGO/KAGRA~\cite{KAGRA:2021kbb}, Taiji~\cite{Ruan:2018tsw,TaijiScientific:2021qgx}, TianQin~\cite{TianQin:2015yph}, Pulsar Timing Array~\cite{Xu_2023} and LISA~\cite{Barausse:2020rsu}.

In this work, we aim to utilize scalar-induced GWs induced by scalar perturbations to explore the early Universe and underlying new physics. 
Despite the successful agreement between the standard model of cosmology and observations, the transition from the inflationary epoch to the standard radiation-dominated (sRD) era remains poorly understood.
Numerous studies suggest the existence of an early matter-dominated (eMD) era following inflation, prior to the sRD era~\cite{Inomata:2019zqy, Lozanov:2022yoy, White:2021hwi, Assadullahi:2009nf, Alabidi:2013lya, Pearce:2023kxp,Inomata:2019ivs,Papanikolaou:2022chm,He:2024luf,Bhaumik:2024qzd}.
In the eMD era, scalar perturbations remain constant even on sub-horizon scales~\cite{Mukhanov:2005sc}.
If the transition from the eMD era to the sRD era is rapid, a sudden change takes place in the equation of state parameter for the Universe, leading to a significant enhancement in the production of scalar-induced GWs~\cite{Inomata:2019zqy, Assadullahi:2009nf, Alabidi:2013lya, Pearce:2023kxp,Inomata:2019ivs}.

The nonlinear interaction of the matter field can result in intriguing physical phenomena, including the formation of solitons and oscillons.~\cite{Rosen:1968mfz,Friedberg:1976me,Coleman:1985ki,Kusenko:1997ad,Amin:2010xe,Kasuya:2002zs,Copeland:1995fq,Broadhead:2005hn,Amin:2010xe,Amin:2011hj,del-Corral:2024vcm,Lozanov:2017hjm,Mahbub:2023faw,Shafi:2024jig}.
The longevity of these compact objects can stem from various reasons. 
Some configurations, such as Q-balls, are stable due to the conservation of topological or nontopological charges~\cite{Rosen:1968mfz,Friedberg:1976me,Coleman:1985ki,Kusenko:1997ad}, while others are sustained by a dynamic equilibrium between attractive forces and dissipative forces, such as oscillons~\cite{Amin:2010xe,Kasuya:2002zs,Copeland:1995fq,Broadhead:2005hn,Amin:2010xe,Amin:2011hj}.
The mass distribution of stable solitons/oscillons is sharply peaked, and they decay rapidly when they reach a critical value.
Therefore, we can treat the transition from the eMD era to the sRD era as an instantaneous process.
For simplicity, we will refer to them collectively as solitons in the following unless otherwise specified.

The separation of solitons increases along with the expansion of the Universe while the physical size remains fixed, which implies that the solitons can be treated as point-like particles for most of the eMD era.
Since the formation of solitons is independent of each other, the number density fluctuations of solitons follow a Poisson distribution, resulting in additional curvature perturbations with a peaked power spectrum upon the nearly scale-invariant one from quantum fluctuations during inflation.
Additionally, the solitons evolve akin to dust gas and can eventually dominate the Universe if their lifetimes significantly exceed the Hubble time.
These induced perturbations of energy density grow at the subhorizon scales during the matter-dominated era, while the amplitude of scalar metric perturbations remains constant. 
The rapid transition from the eMD era to the sRD era also amplifies the GW production, leaving distinct imprints on the GW energy spectrum.
Furthermore, we perform a detailed analysis of the soliton models and the influence on the GW predictions. The key parameters include the soliton lifetime, the initial energy fraction of solitons, and the mean separation of solitons. These free factors of models can affect the duration of the eMD era, the amplitude and the characteristic scale of Poisson-induced perturbations. 
The prediction of the GW energy spectra lies within the sensitivity curves of various GW detectors, where the peak frequency mainly depends on the energy scale of the Universe at the matter-radiation transition.
Additionally, scalar-induced GWs that behave like a dark radiation component, thus contributing to the relativistic degrees of freedom $\Delta N_{\text{eff}}$.
A higher value of $\Delta N_{\text{eff}}$, would delay the radiation-matter equality and change the size of the sound horizon~\cite{PhysRevD.101.063536}, leaving distinct signatures in the cosmic microwave background (CMB) anisotropies~\cite{PhysRevD.87.083008}, baryon acoustic oscillations, and Big Bang nucleosynthesis~\cite{Wallisch:2018rzj,Zhou:2024yke,Chang:2023aba}. 
The latest Planck CMB data constrain $\Delta N_{\text{eff}}<0.3$ at the $95\%$ confidence level~\cite{Planck:2018vyg}. 
Future CMB missions are expected to tighten the constraints on $\Delta N_{\text{eff}}$ by an order of magnitude~\cite{CORE:2016npo}. 
We also constrain the parameter spaces of the soliton models using the upper bounds on scalar-induced GWs inferred from the cosmological constraints on $\Delta N_{\text{eff}}$.

This paper is organized as follows:
In Sec.~\ref{sectiontwo}, we explore the evolution of scalar perturbations in the context of the soliton formation. We calculate the background evolution in the cases where the solitons form during the eMD era and the eRD era, respectively. We then analyze the evolution of the scalar power spectrum in different stages. 
In Sec.~\ref{sectionthree}, we discuss the detailed calculations involved in deriving scalar-induced GWs. 
Section~\ref{result} presents our predictions of GW energy spectra and the constraints on the soliton model based on the observations of $\Delta N_{\text{eff}}$.
We conclude in Sec.~\ref{conclusion}. 

\section{power spectrum of the gravitational potential}
\label{sectiontwo}
\subsection{background dynamics with a eMD era dominated by solitons} 

At the linear order, the perturbed metric in the Newtonian gauge reads
\begin{equation}
\label{ds}
    ds^2=a^2(\eta)\left(-\left(1+2\Phi\right)d\eta^2+\left(\delta_{ij}-2\Phi\delta_{ij}+h_{ij}\right)dx^idx^j\right)\,,
\end{equation}
where the $a(\eta)$ is the scale factor, $\eta$ is the conformal time.
Since we are focusing on the generation of scalar-induced GWs by solitons, in order to simplify our calculations, we have neglected vector perturbations, first-order GWs, and anisotropic stresses. 
This has an ignorable impact on our main conclusions regarding scalar-induced GWs.
According to the Friedmann equations and the continuity equation, in the Universe dominated by a perfect fluid with a general equation of state parameter $\omega$, the scale factor can be expressed as 
\begin{equation}
    a(\eta)=(c_0\eta+c_1)^{\frac{2}{3\omega+1}}\,,
\end{equation}
where $c_0$ and $c_1$ depend on factors such as the energy density of the Universe at the initial stage, the expansion rate, and the evolutionary history of the Universe.

In the early Universe, effective self-resonance resulted in the amplification of perturbations of a real or complex scalar field $\phi$.
As a consequence, the homogeneous scalar condensate fragments into lumps, leading to the formation of solitons, the long-lived, spatially localized structures. 
The solitons are distributed in space by Poisson statistics, with a mean physical separation $d$.
Solitons form rapidly from fragmented scalar field lumps, allowing their formation to be approximated as instantaneous, occurring at time $\eta=\eta_{\text{f}}$.
Before the formation of solitons, the Universe could have been dominated by either matter or radiation. To address this, we consider both scenarios separately.
For convenience, we define the energy density of solitons as $\rho_{\text{s}}$ and the total energy density of the Universe as $\rho_{\text{tot}}$, and we introduce the fractional energy denisity of solitons at $\eta=\eta_{\text{f}}$ as $\Omega_{\text{s,f}}=\langle\rho_{\text{s,f}}\rangle/\rho_{\text{tot,f}}$.
For the solitons (oscillons) formed during the eMD era, due to the relatively low energy density of other surrounding matter at the time of their formation, it can be approximated that the oscillating scalar field (inflation field) transfers almost all of its energy to the solitons. 
This enables the solitons to dominate the energy density within a short period of time. 
In our model, in order to simplify the analysis process, we approximate this process as the solitons immediately dominating the Universe, that is, $\Omega_{s,f} = 1$. 
Conversely, if the solitons are formed during the eRD era, we have $\Omega_{\text{s,f}}\leq1$.
In this paper, we primarily focus on the two cases where $\Omega_{\text{s,f}}=1$ and $\Omega_{\text{s,f}}\ll1$.

After formation, solitons gradually lose energy by emitting relativistic radiation at a slow rate and behave as non-relativistic matter.
The conserved charge of solitons, $Q$, also decreases with time, and solitons rapidly decay once the charge reaches the critical value $Q_{\text{cr}}$, a parameter that depends on the specific soliton model.
For solitons formed during the eRD era, their energy density is negligible at $\eta_{\text{f}}$.
As the Universe expands, the energy density of radiation scales as $\rho_{\text{r}}\propto a^{-4}$ while the energy density of matter scales as $\rho_{\text{m}}\propto a^{-3}$, respectively.
Therefore, for solitons with sufficiently long lifetimes, they could eventually dominate the Universe before decaying.
Since the decay process of solitons is also rapid, we can also assume instantaneous decay.
Therefore, we define the time when the solitons dominate the Universe as $\eta=\eta_{d}$ and the decay time as $\eta=\eta_{r}$. 
The Universe is dominated by matter at $\eta=\eta_{\text{m}}$, and we find the relationship $\eta_{\text{r}}>\eta_{\text{m}}=\eta_{\text{d}}>\eta_{\text{f}}$.

Oscillons can also form in the eMD era. 
The post-inflationary Universe can host a long range of strongly nonlinear processes due to parametric resonance, which is referred to as preheating. The resonance fragments the inflaton condensate and the Universe is strongly inhomogeneous on sub-horizon scales. The energy density is transferred from the oscillating scalar field to the oscillons. Before oscillon formation, the inflaton performs rapid, damped oscillations about a quadratic minimum of the potential after inflation with the frequency much higher than the expansion rate of the Universe.
During the oscillation period, the energy density already scales as $a^{-3}$.
In this scenario, the Universe to be dominated by matter at $\eta=\eta_{\text{m}}$, then solitons form at $\eta=\eta_{\text{f}}$ and dominate the Universe at $\eta_{\text{d}}$, where $\eta_{\text{f}}=\eta_{\text{d}}>\eta_{\text{m}}$, and finally decay at $\eta=\eta_{\text{r}}$.

Similarly, we can find the relationship of the comoving wavenumbers with $k\sim aH$ corresponding to different time nodes: 
the wavenumber at the time of soliton formation, $k_{\text{f}}$; the wavenumber when the Universe begins to be dominated by matter, $k_{\text{m}}$; the wavenumber when solitons start to dominate the Universe, $k_{\text{d}}$; and the wavenumber at the time of soliton decay, $k_{\text{r}}$. 
Additionally, we define the characteristic wavenumber of solitons as $k_{\text{UV}}\equiv\frac{a}{d}$,  beyond which the fluid description of solitons becomes invalid; the average number of solitons within a Hubble event horizon at the time of their formation as $n^3$.
The lifetime of solitons can be described in terms of $N$, where $N$ is defined as $N\equiv\frac{k_{\text{f}}}{k_{\text{r}}}$, and the value of \(N\) depends on the specific soliton model and the physical conditions of the Universe at the time of soliton formation and decay.

Briefly speaking, the extension of the soliton lifespan may, by affecting the distribution of the cosmic energy density, further prolong the duration of the eMD era. At the same time, it will also change the growth time and amplitude of the Poisson-induced perturbations. The initial energy fraction determines the time required from the soliton formation to the domination. 
The average number of solitons and the energy scale of soliton formation together affect the peak wavelength of induced perturbations and induced GWs. 
If solitons were formed during the eRD era, the hierarchy of the relevant scales reads
\begin{equation}
    k_{\text{UV}}>k_{\text{f}}>k_{\text{m}}=k_{\text{d}}>k_{\text{r}}\,,
\end{equation}
and the relationship between them can be expressed through the parameters of the soliton model as follows
\begin{equation}
    \frac{k_{\text{UV}}}{k_{\text{f}}}=n\,,\quad\frac{k_{\text{m}}}{k_{\text{f}}}=2\Omega_{\text{s,f}}\,,\quad\frac{k_{\text{f}}}{k_{\text{r}}}=N\,,
\end{equation}
where the middle equation implies that the matter domination begins when the energy density of radiation and solitons are equal. 

Similarly, if solitons form during the eMD era, we have
\begin{equation}
    k_{\text{UV}}>k_{\text{m}}>k_{\text{f}}=k_{\text{d}}>k_{\text{r}}\,,
\end{equation}
where we assume that $k_{\text{UV}}>k_{\text{m}}$ since the formation of solitons is a sub-horizon process.
The relationship between them at this time is
\begin{equation}
    \frac{k_{\text{UV}}}{k_{\text{f}}}=n\,,\quad\frac{k_{\text{f}}}{k_{\text{r}}}=N\,,\quad\frac{k_{\text{m}}}{k_{\text{f}}}=p\,.
\end{equation}
Here the parameter $p$ is used to describe the ratio between \(k_\text{m}\)  and \(k_\text{f}\) in the case where solitons are formed during the eMD era. Its specific value is determined by the soliton model and the physical conditions of the Universe during the eMD phase. This parameter helps us understand the evolutionary history of the early Universe and facilitates our subsequent calculation of scalar-induced GWs.

We also obtain the evolution of the scale factor $a(\eta)$ and conformal Hubble parameter $\mathcal{H}$ as conformal time $\eta$ in different cases. 
For the solitons formed during the eRD era, the results can be written as 
\begin{equation}
\label{a_eRD}
\begin{split}
    &a_{\text{eRD}}(\eta)=a_{\text{f}}\frac{\eta}{\eta_{\text{f}}}\,,\\
    &a_{\text{eMD}}(\eta)=a_{\text{f}}\frac{\left(\eta+\eta_{\text{m}}\right)^2}{4\eta_{\text{f}}\eta_{\text{m}}}\,,\\
    &a_{\text{sRD}}(\eta)=a_{\text{f}}\frac{\left(\eta_{\text{r}}+\eta_{\text{m}}\right)\left(2\eta-\eta_{\text{r}}+\eta_{\text{m}}\right)}{4\eta_{\text{f}}\eta_{\text{m}}}\,,
\end{split}
\end{equation}
\begin{equation}
    \begin{split}
        &\mathcal{H}_{\text{eRD}}(\eta)=\frac{1}{\eta}\,,\\
        &\mathcal{H}_{\text{eMD}}(\eta)=\frac{2}{\eta+\eta_{\text{m}}}\,,\\
        &\mathcal{H}_{\text{sRD}}(\eta)=\frac{1}{\eta-\frac{\eta_{\text{r}}-\eta_{\text{m}}}{2}}\,.
    \end{split}
\end{equation}
For the solitons formed during the eMD era,
\begin{equation}
\label{a_eMD}
\begin{split}
        &a_{\text{eMD}}(\eta)=a_{\text{f}}\frac{\eta^2}{\eta_{\text{f}}}\,,\\
        &a_{\text{sRD}}(\eta)=a_{\text{f}}\frac{\eta_{\text{r}}\left(2\eta-\eta_{\text{r}}\right)}{\eta^2_{\text{f}}}\,,
\end{split}
\end{equation}
\begin{equation}
\label{H_eMD}
    \begin{split}
        &\mathcal{H}_{\text{eMD}}=\frac{2}{\eta}\,,\\
        &\mathcal{H}_{\text{sRD}}=\frac{2}{2\eta-\eta_{\text{r}}}\,.
    \end{split}
\end{equation}


\subsection{power spectrum of the gravitational potential}\label{power spectrum of the gravitational potential}

Since solitons are rapidly diluted by cosmic expansion after their formation, while their physical size remains unchanged, they become well-separated and can be treated as non-interacting, point-like entities.
Therefore, considering that the solitons are randomly distributed in space according to Poisson statistics with a mean physical separation of $d$, the power spectrum of soliton density perturbations can be expressed as~\cite{Papanikolaou:2020qtd}:

\begin{equation}
\label{delta rho in x}
    \langle\delta\rho_{\text{s}}(\textbf{x})\delta\rho_{\text{s}}(\textbf{x}')\rangle=\frac{4\pi}{3}\left(\frac{d}{a}\right)^3\rho_{\text{s}}^2\delta(\textbf{x}+\textbf{x}')\,,
\end{equation}
where $\delta\rho_{\text{s}}$ represents the density perturbations of $\rho_{\text{s}}$.
The smoothed spectrum of density fluctuations is valid up to $k_{\text{UV}}$. 
For smaller scales, the complex effects of granularity become important and we set $k_{\text{UV}}$ as the cutoff scale.

The Fourier expansion of soliton density perturbations is
\begin{equation}
\label{fourier of delta rho in x}
    \frac{\delta\rho_{\text{s}}(\mathbf{x})}{\rho_{\text{s}}(\mathbf{x})}=\int\frac{d^3\mathbf{k}}{(3\pi)^{3/2}}\delta_{\mathbf{k}}(t)e^{i\mathbf{k}\mathbf{x}}\,.
\end{equation}
By plugging eq.~\eqref{fourier of delta rho in x} into eq.~\eqref{delta rho in x}, the power spectrum of $\delta_{\mathbf{k}}$ can be expressed as
\begin{equation}
    \mathcal{P}_{\delta}(k)=\frac{2}{3\pi}\left(\frac{k}{k_{\text{UV}}}\right)^3\,.
\end{equation}

If solitons form during the eRD era, i.e., $\Omega_{\text{s,f}}\ll1$, 
we define the density contrast of solitons, $\delta_{s}=\frac{\delta\rho_{\text{s}}}{\rho_{\text{s}}}$. 
In this case, isocurvature perturbations are given by
\begin{equation}
    S=\frac{\delta\rho_{\text{s}}}{\rho_{\text{s}}}-\frac{3}{4}\frac{\delta\rho_{\text{r}}}{\rho_{\text{r}}}\approx\frac{\delta\rho_{\text{s}}}{\rho_{\text{s}}}\,,
\end{equation}
where we have used $\Omega_{s}\ll1$. Thus, isocurvature fluctuations can be wiritten as
\begin{equation}
    S_{f}\approx\frac{\delta\rho_{\text{s,f}}}{\rho_{\text{s,f}}}\,,
\end{equation}
and the dimensionless isocurvature power spectrum can then be expressed as
\begin{equation}
    \mathcal{P}_{S}(k)\approx\mathcal{P}_{\delta}(k)=\frac{2}{3\pi}\left(\frac{k}{k_{\text{UV}}}\right)^3\,.
\end{equation}

During the eRD era, such isocurvature perturbations will convert into adiabatic perturbations. 
At the beginning of the sRD era, $\eta=\eta_{\text{r}}$, we can write the Bardeen potential, $\Phi_{k}$, as a summation of the two components,
\begin{equation}
\label{phi}
    \Phi_k(\eta_{\text{r}})=\Phi_{k,\text{inf}}(\eta_{\text{r}})+\Phi_{k,\text{s}}(\eta_{\text{r}})\,,
\end{equation}
where the first component comes from quantum fluctuations during inflation, while the second component is due to the isocurvature perturbations introduced by solitons.
At large scale, curvature perturbations $\zeta$ are related to $\Phi$ as
\begin{equation}
    \langle\Phi_{\textbf{k}}\Phi_{\textbf{k}'}\rangle=\delta(\textbf{k}+\textbf{k}')\frac{2\pi^2}{k^3}\left(\frac{3+3\omega}{5+3\omega}\right)^2\mathcal{P}_{\zeta}(k)\,.
\end{equation}
As predicted by most well-motivated inflationary models, the power spectrum of $\zeta_{k,\text{inf}}$ is approximately scale-invariant,
\begin{equation}
\label{Pphi_inf}
    \mathcal{P}_{\zeta,\text{inf}}=A_{s}\left(\frac{k}{k_*}\right)^{n_{\text{s}}-1}\,.
\end{equation}
From Planck-2018 data, we apply the scalar spectrum amplitude $A_s=2.1\times10^{-9}$, the pivot scale $k_*=0.05$ $\mathrm{Mpc}^{-1}$, and the scalar spectral index  $n_{\text{s}}=0.965$~\cite{Planck:2018jri}.

Then, we investigate the evolution of isocurvature perturbations in the eRD era and eMD era.
At the linear order, scalar perturbations $\Phi$ satisfies
\begin{widetext}
    \begin{equation}
    \label{evolution of phi}
    \Phi''+3\mathcal{H}(1+c_s^2)\Phi'+\left(\mathcal{H}^2\left(1+3c_s^2\right)+2\mathcal{H}'\right)\Phi-c_s^2\Delta\Phi=\frac{a^2\rho_{\text{s}}}{2M_{\text{pl}^2}}c_s^2S\,,
\end{equation}
\end{widetext}
where the prime denotes the deviative with respect to conformal time and $c_s^2$ is the sound speed of the Universe.
Isocurvature perturbations act as a source term in the evolution of scalar perturbations and transform into adiabatic perturbations when the solitons dominate the Universe. 
In the eMD era, $\Phi_{k}$ does not evolve with time for the subhorizon modes.
For $k<k_{\text{s}}$, the mode enters the horizon during the eMD era and becomes constant. 
For $k>k_{\text{s}}$, the mode enter the horizon during the eRD era and decreases until the eMD era begins.
In addition, isocurvature perturbations will convert into adiabatic perturbations and we can obtain the power spectrum $\Phi_{k,\mathrm{s}}$ by solving  eq.~\eqref{evolution of phi}
\begin{equation}
\label{phi}
    \mathcal{P}_{\Phi,\text{s}}(k,\eta_{\text{r}})=\frac{2}{3\pi}\left(\frac{k}{k_{\text{UV}}}\right)^3\left(5+\frac{4}{9}\frac{k^2}{\mathcal{H}_{\text{d}}^2}\right)^{-2}\,.
\end{equation}

If solitons form during the eMD era, i.e., $\Omega_{\text{s,f}}=1$, we can obtain curvature perturbations directly from energy density perturbations of solitons using the relationship
\begin{equation}
    \Delta\Phi-3\mathcal{H}\left(\Phi'+\mathcal{H}\Phi\right)=4\pi Ga^2\delta\rho_{\text{s}}\,.
\end{equation}
Considering the fact that $\Phi$ is a constant during the eMD era, we obtain that
\begin{equation}
    -\frac{2}{3}\frac{k^2}{\mathcal{H}^2_{\text{d}}}\Phi_k-2\Phi_k=\delta_{k}\,.
\end{equation}
where $\Phi_k$ and $\delta_{k}$ are the Fourier forms of $\Phi$ and $\frac{\delta\rho_{\text{s}}}{\rho_{\text{s}}}$.
In this way, we can get the power spectrum of scalar perturbations induced by solitons as
\begin{equation}
\label{Pphi_2}
    \mathcal{P}_{\Phi,\text{s}}(k,\eta_{\text{r}})=\frac{2}{3\pi}\left(\frac{k}{k_{\text{UV}}}\right)^3\left(2+\frac{2}{3}\frac{k^2}{\mathcal{H}_{\text{d}}^2}\right)^{-2}\,.
\end{equation}

Note that if the solitons decay is not a perfectly instantaneous process, the finite duration will affect the modes whose period is larger than the rate of transition.
This gives rise to an additional $k$-dependent suppression factor, which depends on the specific decay behavior of the soliton, in general, can be expressed as~\cite{Inomata:2020yqv}:
\begin{equation}
    \frac{\Phi_{\text{sRD}}}{\Phi^{\text{instant}}_{\text{sRD}}}\approx S^{1/2}\left(\frac{k}{k_{\text{r}}}\right)^{m}\,,
\end{equation}
where $\Phi^{\text{instant}}_{\text{sRD}}$ refers to the instantaneous transition value of scalar perturbations~\cite{Inomata:2020lmk}, $S\sim\mathcal{O}(1)$ and $m$ describes the decay behavior of the soliton which is dependent on the soliton model, and generally satisfies $-1<m<0$\footnote{ 
In certain soliton models, such suppression does not occur, represented effectively by setting $S=1$ and $m=0$.}.

\section{Scalar induced GWs}\label{sectionthree}
In this section, we focus on the energy spectrum of GWs induced  by curvature perturbations we have obtained in the last section.
\subsection{GWs at second order}
Scalar and tensor perturbations of the metric coupled with each other at second order expansion of the Einstein equation.
The equation of motion for tensor perturbations in the Fourier space can be written as
\begin{equation}
    h_{\textbf{k}}''+2\mathcal{H}h_{\textbf{k}}'+k^2h_{\textbf{k}}=4S_{\textbf{k}}\,,
\end{equation}
where $S_{\textbf{k}}$ represents the source term arising from first-order scalar perturbations.
Using the Green's function method, the power spectrum of tensor perturbations can be expressed as
\begin{widetext}
    \begin{equation}
        \mathcal{P}_{h}(\eta,k)=4\int_{0}^{\infty}dv\int_{|1-v|}^{1+v}du\left(\frac{4u^2-\left(1+v^2-u^2\right)^2}{4vu}\right)^2\overline{I^2}(v,u,x)\mathcal{P}_{\zeta}(vk)\mathcal{P}_{\zeta}(uk)\,,
    \end{equation}
\end{widetext}
where the overline denotes the oscillation average and we have defined dimensionless parameter $x=k\eta$. 
$I(v,u,x)$ is the kernel of induced GWs, which takes into account the time evolution of scalar perturbations,
\begin{equation}
    I(v,u,x,x_i)\equiv\int_{x_i}^{x}\frac{a(\Tilde{x})}{a(x)}kG_k(x,\Tilde{x})f(v,u,\Tilde{x},x_i)d\Tilde{x}\,.
\end{equation}
The Green's function $G_{\textbf{k}}(x,\Tilde{x})$ can be obtained by solving the equation
\begin{equation}
\label{Greens function}
    \frac{d^2G_{\textbf{k}}}{dx^2}(x,\Tilde{x})+\left(1-\frac{1}{a(x)}\frac{d^2a(x)}{dx^2}\right)G_{\textbf{k}}(x,\Tilde{x})=\delta(x-\Tilde{x})\,.
\end{equation}

In the following, we define $\Phi_{\textbf{k}}(\eta)\equiv T(k\eta)\phi_{\textbf{k}}$, where $T(k\eta)$ is the transfer function and $\phi_{\textbf{k}}$ is the primordial value.
 Since the velocities of solitons are negligible, we can calculate $f(u,v,\Tilde{x},x_i)$ in terms of the transfer function $T(k\eta)$
 \begin{widetext}
     \begin{equation}
     \label{f}
         f(u,v,\Tilde{x},x_i)=\frac{3}{25(1+\omega)}\left(2(5+3\omega)T(v\Tilde{x})T(u\Tilde{x})+4\mathcal{H}^{-1}\left(T'(u\Tilde{x})T(v\Tilde{x})+T'(v\Tilde{x})T(u\Tilde{x})\right)+4\mathcal{H}^{-2}T'(v\Tilde{x})T'(u\Tilde{x})\right)\,.
     \end{equation}
 \end{widetext}
 
The GW energy spectrum is defined as the fraction of the GW energy density per logarithmic wavelength
\begin{equation}
    \Omega_{\text{GW}}(\eta,k)=\frac{1}{24}\left(\frac{k}{a(\eta)H(\eta)}\right)^2\overline{\mathcal{P}_{h}(\eta,k)}\,.
\end{equation}

The gravitational wave (GW) source term peaks around $\eta_{\mathrm{r}}$ and rapidly diminishes thereafter. Once generated, GWs hardly interact between GW and the background plasma is negligible, and the GW energy density scales with the expansion of the Universe as $a^{-4}$. 
Utilizing entropy conservation, the GW energy spectrum today, $\Omega_{\text{GW},0}$, can be expressed in terms of $\Omega_{\text{GW}}$ in the sRD era as \cite{Inomata:2020tkl,Espinosa_2018}
     \begin{equation}
        \Omega_{\text{GW},0}h^2(k)=0.39\left(\frac{g_c}{106.75}\right)^{-1/3}\Omega_{\text{r},0}h^2\Omega_{\text{GW}}(\eta_c,k)\,,
     \end{equation}
 where $\Omega_{\text{r},0}h^2\sim4.18\times10^{-5}$ is the energy density fraction of radiation evaluated today \cite{Planck:2018vyg}.
 In this paper, we use $g$ to denote the effective relativistic degrees of freedom , and $g_c\equiv g(\eta=\eta_c)$.
 

\subsection{Scalar induced GWs in scenarios of soliton formation}
In the eMD Universe, we can divide the contribution of $I(v,u,x)$ into three components, $I_{\text{eRD}}$, $I_{\text{eMD}}$ and $I_{\text{sRD}}$, which correspond to the GW production in different eras: the eRD era at first, followed by the eMD era and finally the sRD era. 
However, we only focus on GWs generated during the sRD era, because for both $\mathcal{P}_{\Phi_{\text{inf}}}$ and $\mathcal{P}_{\Phi_{\text{s}}}$, the dominant contribution to GWs comes from $I_{\text{sRD}}$, and the GW energy spectrum can be expressed as 
\begin{widetext}
    \begin{equation}
        \Omega_{\text{GW}}(\eta,k)=\frac{1}{6}\int_{0}^{\infty}dv\int_{|1-v|}^{1+v}du\left(\frac{4v^2-\left(1+v^2-u^2\right)}{4uv}\right)^2\overline{\mathcal{I}_{\text{sRD}}^2}(v,u,x,x_{\text{r}})\mathcal{P}_{\zeta}(kv)\mathcal{P}_{\zeta}(ku)\,.
    \end{equation}
\end{widetext}
As mentioned above, the scalar power spectrum $\mathcal{P}_{\Phi}$ can be written as
\begin{equation}
    \mathcal{P}_{\Phi}(k,\eta_{\text{r}})=\mathcal{P}_{\Phi,\text{inf}}(k,\eta_{\text{r}})+\mathcal{P}_{\Phi,\text{s}}(k,\eta_{\text{r}})\,,
\end{equation}
If solitons form during the eRD era, the inflationary scalar perturbation modes, $\Phi_{k,\mathrm{inf}}$, that reenter the horizon during the eRD era, are significantly suppressed so that we can set the cutoff scale of $\Phi_{\text{inf}}$ at $k=k_{\text{m}}$.
If solitons form during the eMD era, primarily referring to oscillons in this paper, $\Phi_{k,\mathrm{inf}}$ with $k>k_{\text{m}}$ correspond to perturbations that remained inside the horizon during inflation, which provide the initial conditions for the self-resonance of inflaton perturbations. Thus, the cutoff condition $k<k_{\text{m}}$ is also valid in this case.
Taking into account that eq.~\eqref{phi} and eq.~\eqref{Pphi_2} are valid only for $k<k_{\text{UV}}$, we can obtain
\begin{equation}
    \mathcal{P}_{\Phi,\text{inf}}(k,\eta_{\text{r}})=\left(\frac{3+3\omega}{5+3\omega}\right)^2A_{s}\left(\frac{k}{k_*}\right)^{n_{\text{s}}-1}\Theta(k_{\text{m}}-k)\,,
\end{equation}
\begin{equation}
    \mathcal{P}_{\Phi,\text{s}}(k,\eta_{\text{r}})=\begin{cases}
        \frac{2}{3\pi}\left(\frac{k}{k_{\text{UV}}}\right)^3\left(5+\frac{4}{9}\frac{k^2}{\mathcal{H}_{\text{d}}^2}\right)^{-2}, &\text{if solitons form during the eRD era};\\
        \frac{2}{3\pi}\left(\frac{k}{k_{\text{UV}}}\right)^3\left(2+\frac{2}{3}\frac{k^2}{\mathcal{H}_{\text{d}}^2}\right)^{-2}, &\text{if solitons form during the eMD era},
    \end{cases}
\end{equation}

Since $\eta_{\text{r}}\gg\eta_{\text{m}}$, whether solitons were produced in the eMD era or the eRD era, the conmoving Hubble parameter $\mathcal{H}$ in the sRD era can be approximately expressed as $\mathcal{H}_{\text{sRD}}\approx\frac{2}{2\eta-\eta_{\text{r}}}$.
We introduce a dimensionless variable $x_{\text{r}}=k\eta_{\text{r}}$, which corresponds to the beginning of the sRD era.
Based on the background evolution given by eq.~\eqref{a_eRD} and eq.~\eqref{a_eMD}, the GW kernel can be obtained by solving the Green's equation eq.~\eqref{Greens function} during the sRD era
\begin{equation}
    I_{\text{sRD}}(v,u,x,x_{\text{r}})=\int_{x_{\text{r}}}^{x}d\Tilde{x}\frac{\Tilde{x}-x_{\text{r}}/2}{x-x_{\text{r}}/2}f(v,u,\Tilde{x},x_{\text{r}})kG_{k}(\Tilde{x},x)\,.
\end{equation}

For simplicity, we define
\begin{equation}
    \mathcal{I}_{\text{sRD}}(v,u,x,x_{\text{r}})=I_{\text{sRD}}(v,u,x,x_{\text{r}})\left(x-\frac{x_{\text{r}}}{2}\right)\,.
\end{equation}

The general solution for the transfer function in the sRD era, preceded by an eMD era can be written as
\begin{equation}
\label{transfer function}
    T(x,x_{\text{r}})=\frac{3\sqrt{3}}{x-x_{\text{r}}/2}\left(A(x_{\text{r}})j_1\left(\frac{x-x_{\text{r}}/2}{\sqrt{3}}\right)+B(x_{\text{r}})y_1\left(\frac{x-x_{\text{r}}/2}{\sqrt{3}}\right)\right)\,,
\end{equation}
where $j_1$ and $y_1$ are the first and second spherical Bessel functions, and $A, B$ are constants which depend on $x_{r}$.
Demanding the continuity of the transfer function and its time dericative at $\eta=\eta_{\text{r}}$, we can determine $A$ and $B$ as
\begin{equation}
    A(x_{\text{r}})=\frac{x_{\text{r}}}{2\sqrt{3}}\sin\left(\frac{x_{\text{r}}}{2\sqrt{3}}\right)-\frac{1}{36}\left(x_{\text{r}}^2-36\right)\cos\left(\frac{x_{\text{r}}}{2\sqrt{3}}\right)\,,
\end{equation}
\begin{equation}
    B(x_{\text{r}})=-\frac{1}{36}(x_{\text{r}}^2-36)\sin\left(\frac{x_{\text{r}}}{2\sqrt{3}}\right)-\frac{x_{\text{r}}}{2\sqrt{3}}\cos\left(\frac{x_{\text{r}}}{2\sqrt{3}}\right)\,.
\end{equation}

In principle, using eq.~\eqref{transfer function} and eq.~\eqref{f}, we can directly obtain the GW energy density $\Omega_{\text{GW}}$.
However, our aim is to establish a relationship between GWs and the parameters of the soliton model, which will further allow us to impose constraints on the soliton model parameters.
Therefore, we seek to derive approximate analytical formulas for induced GWs.
We then obtain the approximate form of $\mathcal{I}_{\text{sRD}}$ on the  scales of $k\gg k_{\text{r}}$ and $k\ll k_{\text{r}}$, and use this approximate form to calculate induced GWs.

In the small-scale approximation, $k\gg k_{r}$ and $x\gg x_{\text{r}}$, the primary contributions to the kernel term $\mathcal{I}_{sRD}$ come from the sum of trigonometric terms.
These include the terms such as
$A(u,v,x_{\text{r}})x_{\text{r}}^4\sin\left(D_{\pm\pm}x_{\text{r}}\right)$, $B(u,v,x_{\text{r}})x_{\text{r}}^4\cos\left(D_{\pm\pm}x_{\text{r}}\right)$, cosine integral terms like $C(u,v,x_{\text{r}})x_{\text{r}}^4\mathrm{Ci}\left(D_{\pm\pm}x_{\text{r}}\right)$, and sine integral terms $D(u,v,x_{\text{r}})x_{\text{r}}^4\mathrm{Si}\left(D_{\pm\pm}\right)$, where $A(u,v,x_{\text{r}})$, $B(u,v,x_{\text{r}})$, $C(u,v,x_{\text{r}})$ and $D(u,v,x_{\text{r}})$ represent the coefficients that depend on $u$, $v$ and $x_{\text{r}}$. Additionally, we define $D_{\pm\pm}=\frac{u\pm v\pm\sqrt{3}}{3\sqrt{3}}$~\cite{Kohri:2018awv}.
Furthermore, we introduce the cosine integral and sine integral functions, defined as:
\begin{equation}
    \mathrm{Ci}(x)=-\int_{x}^{\infty}d\Tilde{x}\frac{\cos\Tilde{x}}{\Tilde{x}}\,,
\end{equation}
\begin{equation}
    \mathrm{Si}(x)=\int_0^xd\Tilde{x}\frac{\sin\Tilde{x}}{\Tilde{x}}\,.
\end{equation}
In this case, we are interested only in the resonant part of induced GWs. Following the calculation of the kernel term of $\mathcal{I}_{\text{sRD}}$, the terms containing $\mathrm{Ci}$ contribute the main part~\cite{Inomata:2019ivs}
\begin{equation}
\label{Ismall}
    \overline{\mathcal{I}_{\text{eRD}}^2}|_{\text{res}}\approx\frac{x_{\text{r}}^8}{5971968}\frac{\left(u^2+v^2-3\right)^4}{u^2v^2}\mathrm{Ci}\left(\left(\sqrt{3}u+\sqrt{3}v-3\right)\frac{x_{\text{r}}}{6}\right)^2\,.
\end{equation}

For the large-scale approximation, considering the cutoff conditions of $\mathcal{P}_{\Phi,\text{inf}}$ and $\mathcal{P}_{\Phi,\text{s}}$, the integrals over $u$ and $v$ are primarily dominated by the region where $u+v-1$ is large and $u-v$ is small.
Under these approximations, we find that
\begin{equation}
\label{Ilarge}
    \overline{\mathcal{I}^{2}_{\text{sRD}}}|_{\text{lar}}\approx\frac{(u+v-1)^4x_{\text{r}}^8}{9102222}\left(4\mathrm{Ci}\left(\frac{x_{\text{r}}}{2}\right)^2+\left(\pi-2\mathrm{Si}\left(\frac{x_{\text{r}}}{2}\right)\right)^2\right)\,.
\end{equation}

Therefore, the energy spectrum of induced GWs can be expressed as
\begin{equation}
\label{omega}
    \Omega_{\text{GW}}=\Omega_{\text{GW,inf,res}}+\Omega_{\text{GW,inf,lar}}+\Omega_{\text{GW,s,res}}+\Omega_{\text{GW,inf,lar}}\,.
\end{equation}
These four terms represent the contribution of $\mathcal{P}_{\Phi,\text{inf}}$ and $\mathcal{P}_{\Phi,\text{s}}$ on small and large scales, respectively.
Using eq.~\eqref{Ismall} and eq.~\eqref{Ilarge}, we can derive the following expression
\begin{equation}
\begin{split}
    \Omega_{\text{GW,inf,lar}}=&1.14\times10^{-9}\left(4\mathrm{Ci}\left(\frac{x_{\text{r}}}{2}\right)^2+\left(\pi-2\mathrm{Si}\left(\frac{x_{\text{r}}}{2}\right)\right)^2\right)S^2A_s^2x_{\text{r}}^{8}\left(\frac{k}{k_{\text{r}}}\right)^{4m}\left(\frac{k}{k_*}\right)^{2n_s-2}\\
    &\int dv\int du\left(4v^2-(1+v^2-u^2)^2\right)^2(u+v-1)^4(uv)^{2m+n_s-3}\,,
\end{split}
\end{equation}
\begin{equation}
\begin{split}
    \Omega_{\text{GW,s,lar}}=&2.6\times10^{-10}\left(4\mathrm{Ci}\left(\frac{x_{\text{r}}}{2}\right)^2+\left(\pi-2\mathrm{Si}\left(\frac{x_{\text{r}}}{2}\right)\right)^2\right)S^2x_{\text{r}}^{8}\left(\frac{k}{k_{\text{r}}}\right)^{4m}\left(\frac{k}{k_{\text{UV}}}\right)^{6}\\
    &\int dv\int du\left(4v^2-(1+v^2-u^2)^2\right)^2(u+v-1)^4(uv)^{2m+1}\frac{1}{\left(a+b\frac{k^2v^2}{\mathcal{H}^2_{\text{d}}}\right)^2\left(a+b\frac{k^2v^2}{\mathcal{H}^2_{\text{d}}}\right)^2}\,,
\end{split}
\end{equation}
\begin{equation}
    \Omega_{\text{GW,inf,res}}=1.4\times10^{-6}\left(\frac{1}{4}\right)^{2m+n_s}S^2A_s^2x_{\text{r}}^7\left(\frac{k}{k_{\text{r}}}\right)^{4m}\left(\frac{k}{k_*}\right)^{2n_s-2}\int_{s_{\text{m}}}^{s_{\text{m}}}\left(1-s^2\right)^2\left(3-s^2\right)^{2m+n_s-1}\,,
\end{equation}
\begin{equation}
    \Omega_{\text{GW,s,res}}=3.2\times10^{-7}\frac{S^2}{b^2}\left(\frac{1}{4}\right)^{2m-2}x_{\text{r}}^7\left(\frac{k}{k_{\text{r}}}\right)^{4m}\left(\frac{k}{k_{\text{UV}}}\right)^6\left(\frac{\mathcal{H}_d}{k}\right)^8\int_{-s_{\text{UV}}}^{s_{\text{UV}}}ds\left(1-s^2\right)^2\left(3-s^2\right)^{2m-1}\,.
\end{equation}
Here, the lower and upper limits of the integral are described by $s_{\text{m}}$ and $s_{\text{UV}}$, which come from the cutoff condition of $\Theta(k_{\text{m}}-k)$ and $\Theta(k_{\text{UV}}-k)$.
The upper limit $s_{\text{UV}}$ is a function of $k$
 \begin{equation}
     s_{\text{max}}=\begin{cases}
         1,&\frac{k}{k_{\text{max}}}\leq\frac{2}{1+\sqrt{3}}\,;\\
         2\frac{k_{\text{max}}}{k}-\sqrt{3},&\frac{2}{1+\sqrt{3}}<\frac{k}{k_{\text{max}}}\leq\frac{2}{\sqrt{3}}\,;\\
         0,&\frac{k}{k_{\text{max}}}>\frac{2}{\sqrt{3}}\,.
     \end{cases}
 \end{equation}
 
The result of eq.~\eqref{omega} also contains contributions from the cross terms between $\mathcal{P}_{\Phi,\text{inf}}$ and $\mathcal{P}_{\Phi,\text{s}}$. 
In most cases either $\mathcal{P}_{\Phi,\text{inf}}$ or $\mathcal{P}_{\Phi,\text{s}}$ dominates the total $\mathcal{P}_{\Phi}$, $\Omega_{\mathrm{GW}}$ is mainly contributed from the dominant term and the cross term is negligible.

\section{results and the constraints to soliton model}\label{result}
\begin{figure}
\centering
    \begin{minipage}[t]{0.45\linewidth}
        \includegraphics[width=\linewidth]{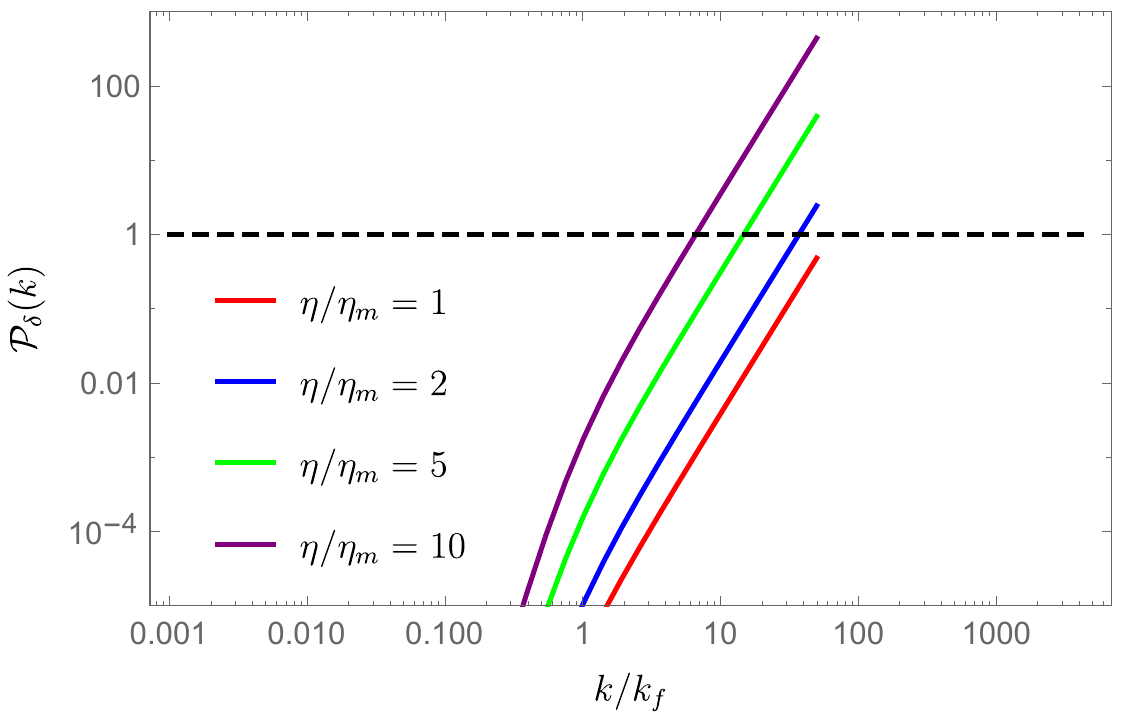}
    \end{minipage}
\hfill
    \begin{minipage}[t]{0.45\linewidth}
        \includegraphics[width=\linewidth]{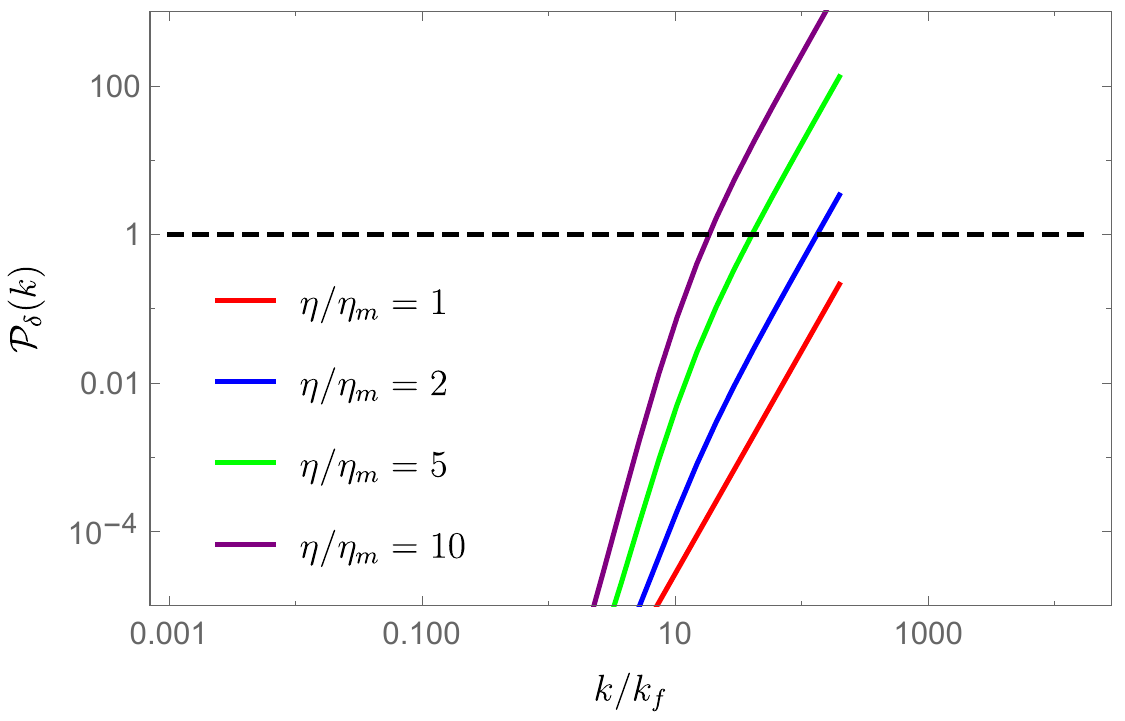}
    \end{minipage}
    
    \caption{This figure illustrates the temporal evolution of soliton density perturbations during the eMD era. 
    The region above the black dashed line represents the scale at which these perturbations reach nonlinearity, providing a conservative estimate for our calculations. 
    The left panel shows the scenario where solitons form during the eRD era, while the right panel illustrates their formation during the eMD era.}
    \label{figcut}
\end{figure}

\begin{figure}
\centering
    \begin{minipage}[t]{0.45\linewidth}
        \includegraphics[width=\linewidth]{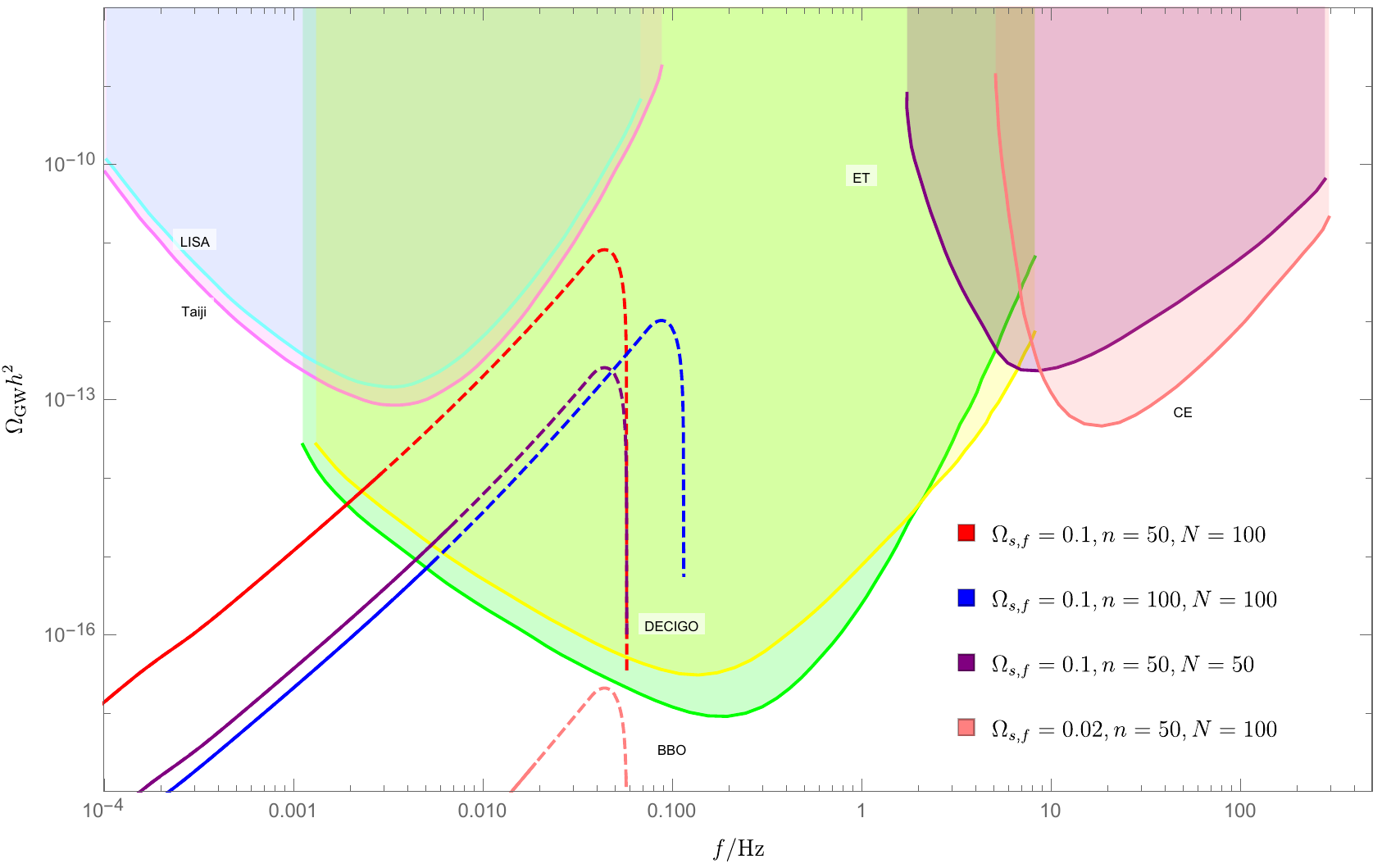}
    \end{minipage}
\hfill
    \begin{minipage}[t]{0.45\linewidth}
        \includegraphics[width=\linewidth]{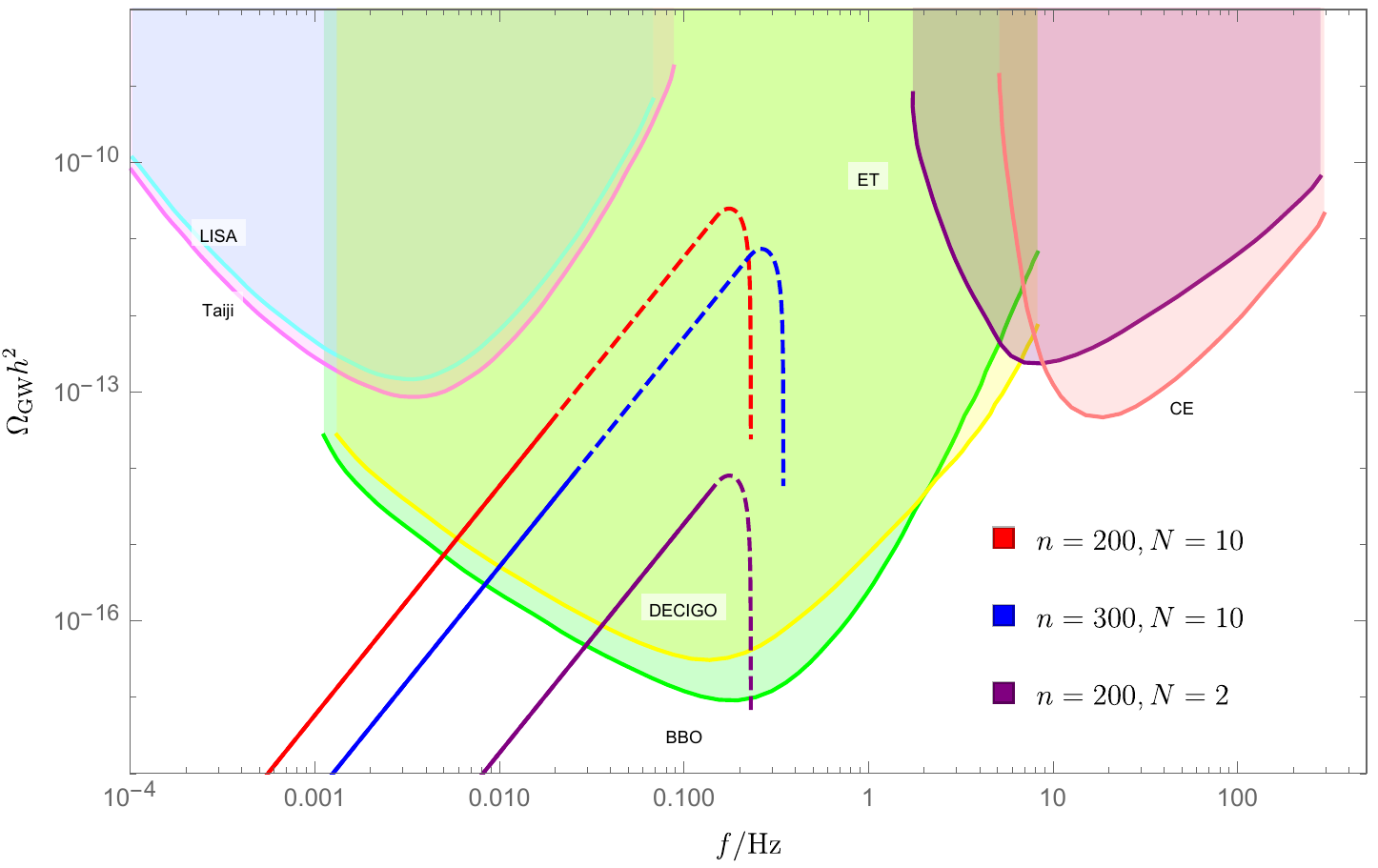}
    \end{minipage}
    
    \caption{This figure shows the power spectra of scalar-induced GWs under various parameter choices for the soliton model and we set $k_{\text{f}}=0.001\mathrm{Hz}$. 
    The left panel represents the scenario where solitons form during the eRD era, while the right panel illustrates soliton formation during the eMD era. 
    The dashed line indicates the scale at which soliton-induced density perturbations reach nonlinearity, making the present results conservative estimates for this portion of the induced GWs.
    For simplicity, we set $S=1$ and $m=-\frac{1}{2}$.}
    \label{figgw}
\end{figure}

\begin{figure}
\centering
    \begin{minipage}[t]{0.45\linewidth}
        \includegraphics[width=\linewidth]{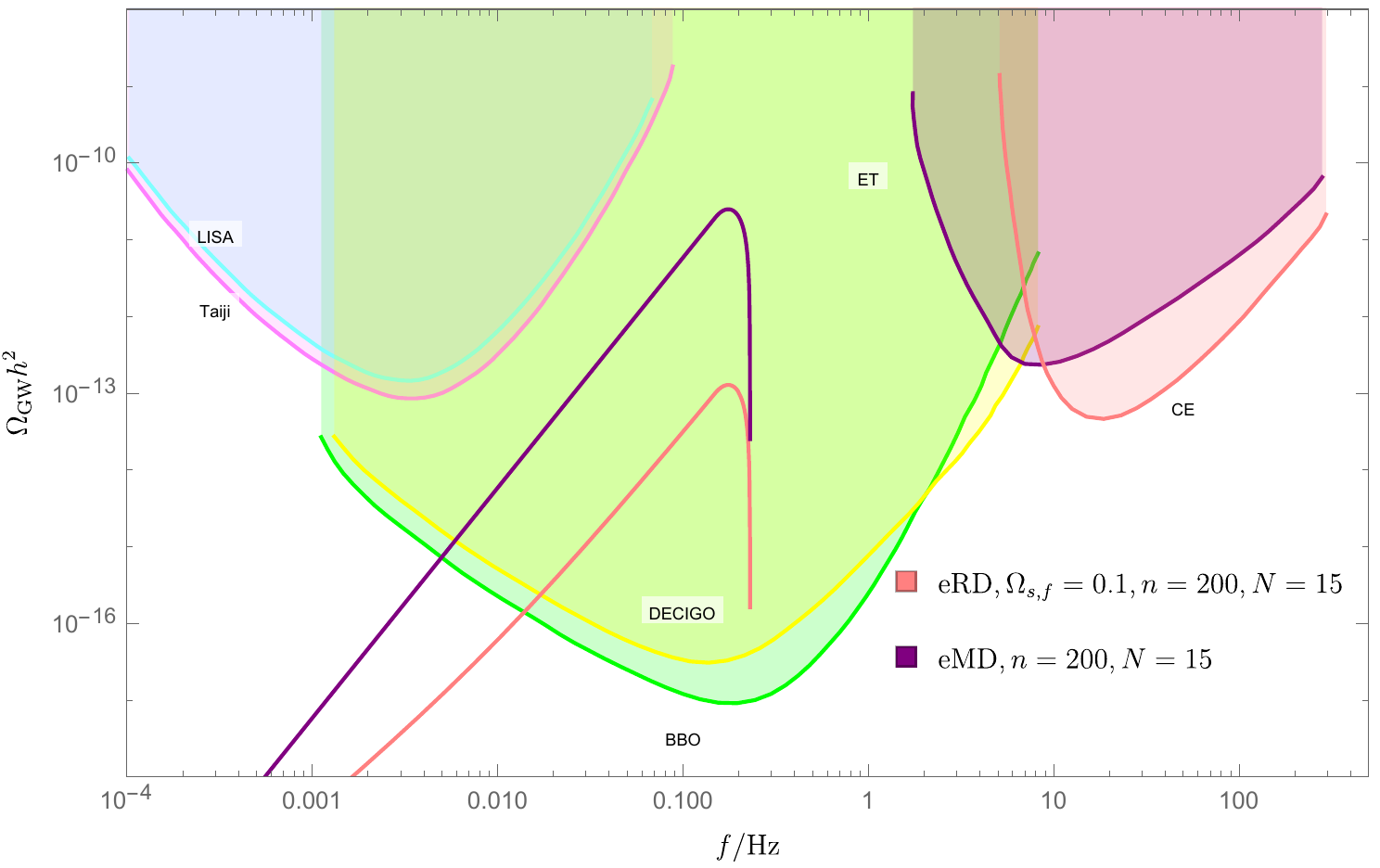}
    \end{minipage}
\hfill
    \begin{minipage}[t]{0.45\linewidth}
        \includegraphics[width=\linewidth]{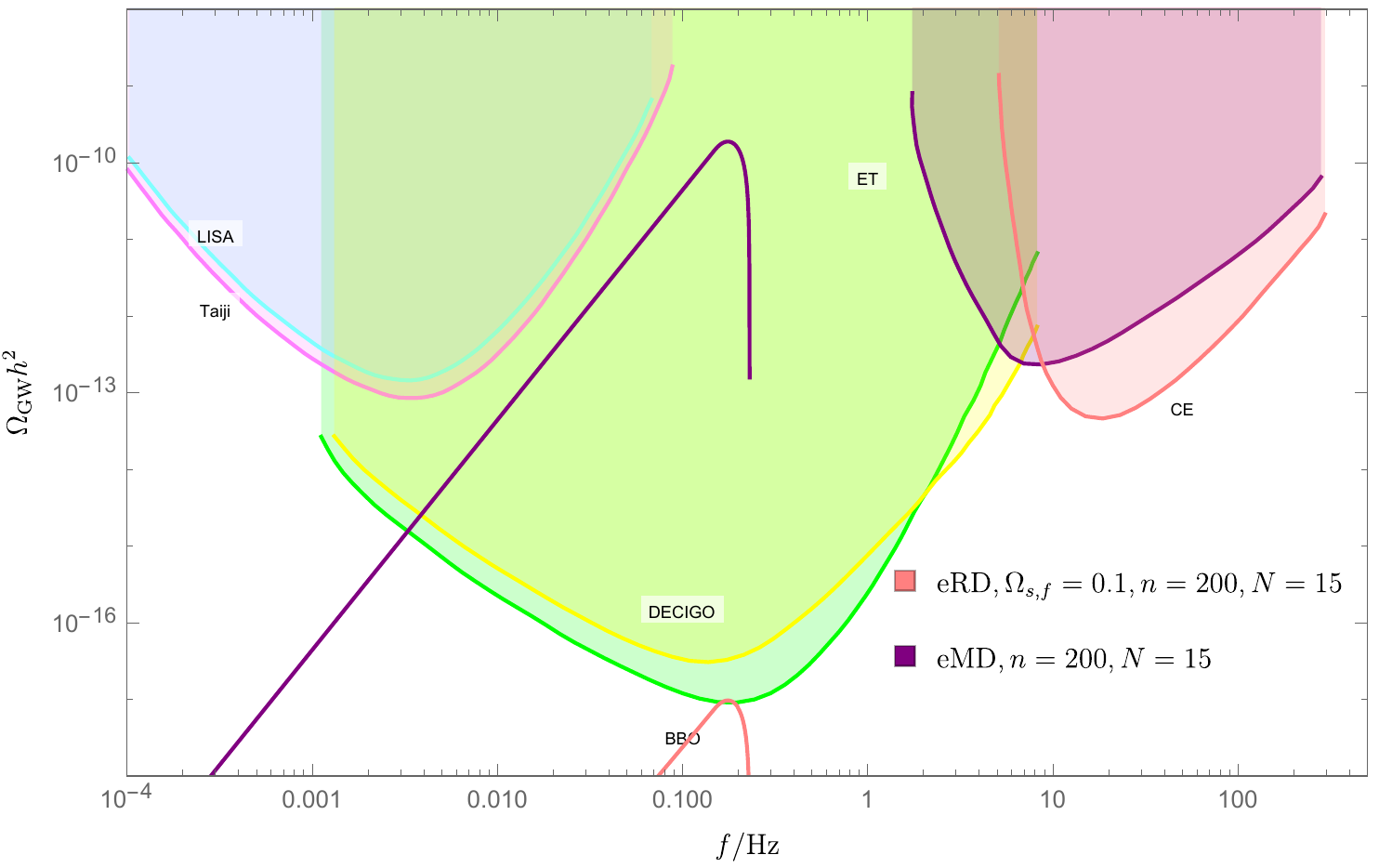}
    \end{minipage}
    
    \caption{This figure illustrates differences in the GW energy spectra induced by solitons forming during the eMD era and the eRD era, where we keep other parameters the same and set $k_{\text{f}}=0.001\mathrm{Hz}$. 
    The pink solid line represents soliton formation during the eRD era, while the purple solid line corresponds to soliton formation during the eMD era. 
    In the left panel, both scenarios assume an identical soliton formation number density and an equal duration of the soliton-dominated eMD era. 
    In the right panel, both scenarios assume the same soliton formation number density and the same soliton lifetime.
    For simplicity, we set $S=1$ and $m=-\frac{1}{2}$.}
    \label{figto}
\end{figure}

\begin{figure}
\centering
    \begin{minipage}[t]{0.45\linewidth}
        \includegraphics[width=\linewidth]{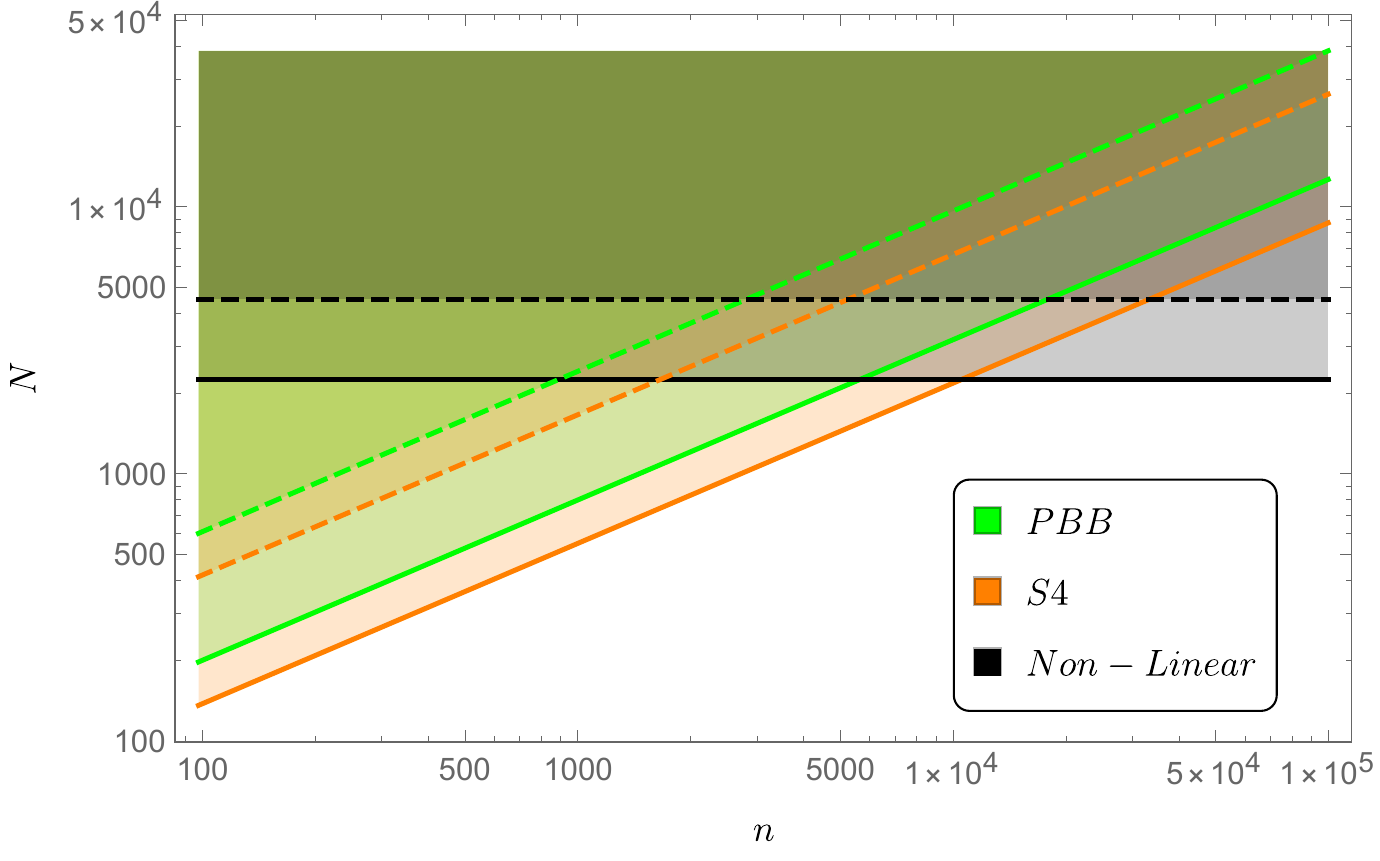}
    \end{minipage}
\hfill
    \begin{minipage}[t]{0.45\linewidth}
        \includegraphics[width=\linewidth]{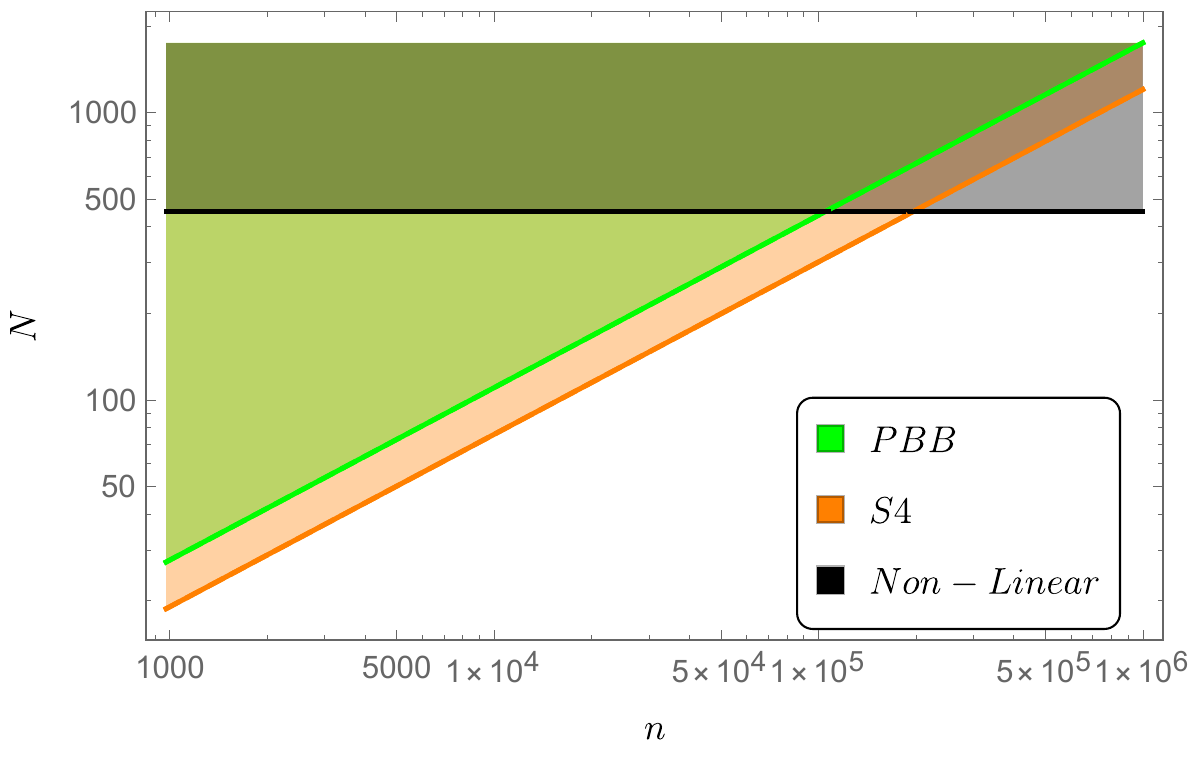}
    \end{minipage}
    
    \caption{This figure illustrates the constraints imposed by induced GWs on the soliton model. 
    The orange curve denotes the constraint applied by PBB, while the yellow curve represents the constraint set by $\mathrm{S4}$ on the soliton model. 
    The black curve marks the threshold at which primordial density perturbations become nonlinear, beyond which our constraints no longer hold. 
    In the left panel, we present the scenario where solitons form during the eRD: the solid line corresponds to the case of $\Omega_{\text{s,f}}=0.1$, and the dashed line to $\Omega_{\text{s,f}}=0.05$. 
    The right panel shows the scenario where solitons form during the eMD era.
    For simplicity, we set $S=1$ and $m=-\frac{1}{2}$.}
    \label{figconstrain}
\end{figure}
Before presenting our results, we first outline the relevant scales to which our discussion applies. 
During the eMD era, the near-zero pressure of the Universe leads to a growth in small-scale density perturbations over time. 
If this period of matter dominance extends too long, density perturbations will inevitably become nonlinear, and our conclusions will no longer be applicable. 
This limitation has been discussed in prior research, where the nonlinear scale of primordial density perturbations is denoted by $k_{\text{nl}}\sim\mathcal{P}_{\zeta}^{-1/4}\mathcal{H}$. 
However, for density perturbations induced by solitons, small-scale perturbations are significantly larger than primordial density perturbations, necessitating a careful assessment of the scales over which our conclusions remain valid. 
As shown in Fig.~\ref{figcut}, soliton-induced density perturbations at small scales can reach nonlinearity relatively quickly during the eMD era compared to primordial density perturbations. 
At nonlinear scales, solitons interact to form small-scale structures, and on these scales, the gravitational wave power spectrum follows $\Omega_{\text{GW}}\sim k^{3/2}$\cite{Fernandez:2023ddy}. 
Since all of our previous calculations are based on the linear perturbation theory, thus, our results represent a conservative estimate of $\Omega_{\mathrm{GW}}$.

A main finding of this paper is that even with a brief eMD era, a soliton-dominated early Universe can still produce sufficiently strong induced GWs.
In Fig.~\ref{figgw}, we present  scalar-induced GWs generated with different soliton model parameters. 
Distinguished from previous studies, we do not require a long duration of the eMD era to amplify primordial scalar perturbations as efficient GW sources. The duration of the eMD era satisfies the condition $\frac{k_{\text{m}}}{\eta_{\text{r}}}\sim\mathcal{O}(100)$. 
Since the number of solitons within a Hubble horizon at the formation time spans in a large range, depending solely on the soliton model, Poisson-induced density perturbations can increase to the $\mathcal{O}(1)$ level and produce strong GWs within the short duration of the eMD era.

We also consider the case that solitons may form during the eMD era. This scenario introduces several key differences compared with other works discussing the generation of GWs by solitons~\cite{Lozanov:2023aez}:
First, if the soliton forms during the eRD era, its density perturbation is expected to initially behave as isocurvature perturbations. 
As the energy density of the soliton increases with time, the isocurvature perturbations gradually transition to the adiabatic perturbations. 
However, if the soliton forms during the eMD era, the physical situation differs. 
In this case, an oscillating scalar field dominates the Universe, and solitons emerge through resonant fragmentation of the field. 
As a result, density perturbations of solitons behave as adiabatic perturbations from the moment of its formation. 
Since solitons capture most of the energy of scalar field, they immediately dominate the Universe upon formation.
Second, if solitons form during the eMD era, they will directly govern the dynamics of the Universe. In this case, the lifetime of the solitons determines the duration of the eMD era. 
Thus, even with a very short soliton lifetime, $N\sim 1$, strong GWs can still be induced, as shown in Fig.~\ref{figgw}.
Finally, as shown in eq.~\eqref{a_eRD} and eq.~\eqref{a_eMD}, the expansion rates of the Universe during the matter-dominated and radiation-dominated eras are different. 
This leads to the conclusion that, in the two scenarios where solitons form during the eMD era and the eRD era, respectively, the induced GWs will be stronger in the former scenario, even when both cases share the same soliton model parameters as shown in Fig.~\ref{figto}.
This is primarily due to two reasons. 
First, during the eMD era, the Universe expands more slowly, which enlarge the ratio between the peak wavelength of induced perturbations and the Hubble radius at the decay time. 
Second, when solitons form during the eRD era, they require additional time to gradually become dominant.

We also differentiate our work from other studies that consider deviations of the cosmic equation of state from the standard radiation-dominated value, $\omega=\frac{1}{3}$, before the eMD era~\cite{Domenech:2024wao,Bhaumik:2024qzd}. 
These studies indicate that a smaller equation of the state parameter \( \omega \) tends to suppress the amplitude of the induced GWs, which may appear to be in tension with our conclusions. 
However, this apparent discrepancy arises because, for smaller \( \omega \), it takes a longer time for dust to dominate the Universe, thereby shortening the duration of the eMD era. 
In contrast, solitons formed during the eMD era can quickly become the dominant component, and their number density perturbations are adiabatic rather than isocurvature.

The peak of the induced gravitational wave power spectrum is located at $k\sim k_{\text{UV}}$, which in turn depends on the energy scale at the time of soliton formation and the number density at that time.
Observational constraints on the early Universe require that the presence of the eMD era does not adversely affect primordial nucleosynthesis. 
Therefore, in the scenario we consider, solitons are assumed to form at an energy scale $T>10^2\mathrm{GeV}$ in order to prevent the affects on primordial nucleosynthesis. 
Under these conditions, the induced GWs could potentially be detected by future GW observatories, as illustrated in Fig.~\ref{figgw} where we assume $k_{\text{f}}=0.001\mathrm{Hz}$, We choose this parameter set to demonstrate the generation and characteristics of GWs. 
The generated gravitational wave signals can also be detected by the Pulsar Timing Array~(PTA) for the soliton decay at a lower energy scale. Future research can further explore the detailed relationship between the soliton model and the detectability of the induced GWs by PTA.

Many studies have investigated scenarios in which the early Universe is dominated by components that follow a Poisson distribution, such as PBHs~\cite{Bhaumik:2024qzd,Domenech:2024wao,Bhaumik:2022pil,Domenech:2024wao,Domenech:2020ssp,Papanikolaou:2020qtd,Bhaumik:2020dor}. 
However, when extending the discussion to PBH models, parameters such as the PBH lifetime, formation time, initial energy fraction, and average spacing are not mutually independent but must satisfy various constraints. 
As a result, our conclusions cannot be directly applied to PBH scenarios.
First, PBHs can only form during the eRD era, and a finite amount of time is required for them to come to dominate the Universe. In contrast, solitons can form during eMD era and rapidly become the dominant component.  
Second, soliton formation is a sub-horizon process, whereas PBHs typically form on super-horizon scales. 
Therefore, even if solitons are formed during the eRD era, the peak of the induced GW spectrum appears at smaller scales compared to that of PBHs. 
Finally, the decay dynamics of solitons and PBHs differ and are characterized by the parameters \( m \) and \( s \). 
For PBHs, \( m = -1/3 \), whereas for solitons, \( m \) takes different values depending on the specific soliton model. In general, the allowed range for solitons is \( -1 < m < 0 \). 
Qualitatively, a larger value of \( m \) corresponds to a faster transition from the eMD era to sRD era, leading to stronger GW signals. 
Although \( m \) lies within the range \( -1 < m < 0 \), its influence on the peak amplitude of the GW spectrum appears as a factor of \( (k_{\text{UV}}/k_{\text{r}})^{4m} \), which can result in an enhancement exceeding \( \mathcal{O}(1) \). 

In this section, we also discuss the constraints on soliton models based on the results we obtained.
We discuss a general soliton model and parameterize the model in terms of $N$, $\Omega_{\text{s,f}}$ and $n$, which we introduced in section~\ref{sectiontwo}.
From the cosmological point of view, GWs constitute of dark radiation and can be parameterised by an correction of the effective degree of freedom, $\Delta N_{\text{eff}}=N_{\text{eff}}-N^{\text{SM}}_{\text{eff}}$.
A higher $N_{\text{eff}}$ can delay radiation-to-matter equality and change the size of the sound horizon, which can leave features on CMB anisotropies, baryon acoustic oscillations and Big-Bang nucleosynthesis. In this way, we can get the GW upper bounds inferred from cosmological constraints on $N_{\text{eff}}$.
The limit given by the current and future observations on $\Delta N_{\text{eff}}$ is~\cite{Cang:2022jyc}
\begin{equation}
        \Delta N_{\text{eff}}=\begin{cases}
            0.175, &Planck+\text{BAO}+\text{BBN}\,\text{(PBB)}\,,\\
            0.027, &\text{CMB Stage IV}\,\text{(S4)}\,.
        \end{cases}
\end{equation}
This gives the upper bound on GWs density at $95\%$ C.L.
\begin{equation}
\label{gwmax}
        \Omega_{\text{GW,bou}}=\begin{cases}
            2.11\times10^{-6}, &\text{PBB}\,,\\
            3.25\times10^{-7}, &\text{S4}\,.
        \end{cases}
\end{equation}
It should be noted that if one wants to impose restrictions on a specific soliton model, other observational constraints on the model are required. Here, we only focus on the enhancement phenomenon of the induced GWs during the period when solitons dominate in the early Universe, and on how to use induced GWs to impose constraints on the general soliton model.
Since we are primarily interested in the peak value of the GWs, we focus only on the resonant component of the GWs. Utilizing eq.~\eqref{gwmax} we can derive the constraints on the parameter space of the soliton model as follows:
\begin{equation}
    \begin{cases}
        n^{4m-1}N^{7+4m}\Omega^8_{\text{s,f}}<1.515\times 10^{3}\,S^{-2}\Omega_{\text{GW,bou}}, &\text{If solitons form during the eRD era.}\,,\\
        \,\,\,\,\,\,\,\,\,n^{4m-1}N^{7+4m}<7.634\times 10^{3}\,S^{-2}\Omega_{\text{GW,bou}}, &\text{If solitons form during the eRD era.}\,.  
    \end{cases}
\end{equation}

Fig.~\ref{figconstrain} illustrates the constraints imposed on the parameter space of the soliton model from $\Delta N_{\text{eff}}$. 
The models with large separation and long lifetime of solitions can be ruled out for the overproduction of GWs.
The constraints on the soliton models become weaker for solitons formed in the eRD era because of the smaller peak value of $\Omega_{\text{s,f}}$.

Note that if the soliton lifetime is too long, the density perturbations during the eMD era will grow over time and eventually become nonlinear. 
As a result, we restrict our discussion to the case where $\frac{k}{\eta_{\text{r}}}<450$.

\section{Conclusions} \label{conclusion}
In this paper, we explore the scenario of an eMD era in the Universe driven by solitons and the generation of scalar-induced GWs.
Distinguished from previous studies, we both consider soliton formation during both the eRD and eMD eras.
In the scenario where solitons form during the eRD era, the Universe is initially radiation-dominated. 
Solitons then gradually form and start to dominate the Universe, marking the beginning of eMD. 
Solitons may also form during the eMD era, in which case the Universe is initially dominated by an oscillating scalar field. 
Since the solitons contain the majority of the energy of scalar field, they can be considered to dominate the Universe as soon as they form. 
Since the solitons formation is a random process, their spatial distribution generally follows a Poisson distribution, resulting in an induced power spectrum of density perturbations with an ultraviolet cutoff at a scale related to the average soliton separation, $k_{\text{UV}}$. 
For solitons form during the eRD era, the Poisson distribution induces isocurvature perturbations that transit into curvature perturbations once solitons dominate the Universe. 
For solitons form in the eMD era, density perturbations of solitons in this case are adiabatic from the formation time. 
Density perturbations that originate both from inflation and the Poisson distribution increase during the soliton-dominated era and the corresponding GW sources reach the maximum at around the decay time of solitons, predicting characteristic GW energy spectra that are expected to be observed by multiband GW observers. We also impose new restrictions on the soliton models that overproduce GWs from the upper bound of $N_{\mathrm{eff}}$.



In this work, our calculations assume small density perturbations remain in the linear regime, $\frac{\delta\rho}{\rho}\lesssim 1$. However, energy density gradually increases over time during the eMD era and may readily reach nonlinear threshold for long-lived solitons. The solitons might also experience colliding with each other in the non-linear regime. In this paper, we present a conservative analysis without considering these nonlinear interactions. 
These effects would require a more detailed analysis, which we leave for future work.

\begin{acknowledgments}
    This work is supported in part by the National Key Research and Development Program of China Grants No. 2020YFC2201501 and No. 2021YFC2203002, in part by the National Natural Science Foundation of China Grants No. 12105060, No. 12147103, No. 12235019, No. 12075297 and No. 12147103, in part by the Science Research Grants from the China Manned Space Project with NO. CMS-CSST-2021-B01, in part by the Fundamental Research Funds for the Central Universities.
\end{acknowledgments}



\bibliography{citeLib}

\end{document}